\documentclass[a4paper,11pt]{article}

\usepackage{amsmath}
\usepackage{amssymb}
\usepackage{indentfirst}
\usepackage{graphicx}
\usepackage[utf8]{inputenc}
\usepackage[dvipsnames]{xcolor}
\usepackage{caption}
\usepackage{tikz}
\usetikzlibrary{arrows.meta,positioning,shapes.geometric,fit,calc}

\usepackage{jcappub}
\hypersetup{colorlinks=true, linkcolor=blue, urlcolor=blue, citecolor=blue, linktocpage=true}

\begin{document}

\title{Learning physical event representations by incorporating neural fields into autoencoders}

\author[1,2]{R.~Fitagdinov,}
\author[1,2]{I.~Kharuk}

\affiliation[1]{Institute for Nuclear Research of the Russian Academy of Sciences, \\ 60th October Anniversary Prospect 7a, Moscow, 117312, Russia}
\affiliation[2]{Moscow Institute of Physics and Technology,\\ Institutsky lane, 9, Dolgoprudny, Moscow region, 141700, Russia}

\emailAdd{fitagdinov.rr@phystech.edu, ivan.kharuk@phystech.edu}

\abstract{
Autoencoders are often used to get a compact representation of the data, this enabling the analysis of its structure.
For physical experiments with spatially distributed detectors, such as Telescope Array or IceCube, this entangles physical properties of the event and detector geometry.
We propose a geometry-conditioned autoencoder whose encoder infers physical event representations from measurements and whose shared multilayer perceptron decoder learns the physical response as a neural-field surrogate.
This supports target-agnostic, unsupervised anomaly searches, with decoder conditioning controlling the contribution of selected physical parameters to the anomaly criterion.
We demonstrate the framework using simulated Telescope Array surface-detector data as a case study.
The model predicts station responses more accurately than standard analytical reconstruction and recovers withheld responses from sparse measurements, enabling super-resolution.
Beyond these applications, we discuss how the suggested geometry-conditioned autoencoder can be used to reduce the domain shift via calibration, to estimate event parameters, and for fast Monte Carlo simulations.
}

\maketitle

\section{Introduction}
\label{sec:INTRO}

Modern physics experiments, such as the LHC, Baikal-GVD, and Telescope Array, generate large volumes of multidimensional data~\cite{Evans2008,Avrorin2019,AbuZayyad2012}.
Their analysis requires both efficient processing and an understanding of the underlying physical processes.
Rare event classes may reveal physical processes that are hidden beneath the dominant background.
For example, decays of super-heavy dark matter ~\cite{Berezinsky:1997hy} may produce unusual air shower structure differing from the dominant hadronic background.
Unexpected events may also expose limitations of the assumed interaction or source models.
Searches optimized for a specified signal can miss phenomena with different, unanticipated signatures.

Autoencoders provide compact event representations from which the detector measurements can be reconstructed~\cite{Finke2021,Ostdiek2021}.
The resulting latent space can be used to search for rare or atypical events that do not require a detailed model of the anomalous signal~\cite{Finke2021,Kasieczka2021}.
Common criteria include a large reconstruction error or a \(\sigma\)-based cut relative to a reference distribution.

Such criteria can respond to detector effects rather than the unexpected phenomena being sought since latent representation can mix event physics with the configuration of the detector.
Known physical parameters, such as energy and zenith angle, can also enter the representation.
However, their influence is not always relevant to the anomaly under study.
For example, a search may target unusual shower development rather than an unusual primary-particle energy.
In this case, it is desirable to reduce the influence of energy on the anomaly score~\cite{Louppe2017,Shimmin2017}.
The problem is therefore not only to compress an event, but to choose which aspects of the event the representation should describe.

The distinction between event properties and recording conditions is particularly important for spatially distributed detectors, such as Telescope Array, Pierre Auger Observatory, or IceCube.
A physical event can be sampled by different sets of detector stations, producing different collections of measurements.
These observations should be related through a common description of the event.
Station positions are needed to predict each measurement, but should not define the event itself.
This motivates an event representation in which detector geometry is supplied separately.

We address this problem with a geometry-conditioned autoencoder.
The encoder compresses the available detector measurements into a common latent vector.
A shared multilayer perceptron (MLP) uses this vector and the coordinates of a detector element to predict its response.
Its outputs describe local observables, such as signal amplitudes or arrival times, and can be queried at all station positions to get the full detector response.
The latent vector describes the event physics, while the coordinates specify where its response is evaluated.
The MLP decoder thus defines a continuous response field over spatial coordinates conditioned on the latent vector.
Thus, it is a latent-conditioned neural field, built using an established coordinate-based representation~\cite{Mildenhall2020}.
In air-shower analysis applications, this field acts as a learned lateral distribution and shower-front model.

Beyond making the geometry explicit, we also seek control over which physical variations define an anomaly.
For this purpose, selected physical parameters enter the decoder through a separate input.
We combine this conditioning with established adversarial representation learning~\cite{Louppe2017,Shimmin2017,Wang2018Invariant}.
A critic attempts to recover these parameters from the latent vector.
The encoder is trained to oppose this prediction while preserving the information needed for reconstruction.
Conditioning supplies the decoder with the known physical parameters, while the critic discourages their redundant encoding in the latent vector.
For energy conditioning, the intended question is whether a shower has unusual structure for its known energy.

We demonstrate the framework using Monte Carlo (MC) simulated data from the Telescope Array (TA) surface detector~\cite{AbuZayyad2012}.
TA records extensive air showers' footprints on ground level with an array of 507 scintillation stations.
As a practical application, we consider photon-induced showers as a controlled anomaly against a dominant proton background.
The autoencoder is trained on a mixed sample with a photon fraction below \(1\%\).
We pose the problem as a general anomaly search --- finding atypical events on top of dominant proton-induced air showers. 
Primary-particle labels are used only to evaluate the sensitivity of the search, unlike the supervised classification, which requires labeled examples of the signal being sought~\cite{Finke2021,Kasieczka2021}.

The same encoder--decoder connection also supports super-resolution.
The encoder receives measurements from a thinned station grid.
The decoder reconstructs the responses of the original event, including stations withheld from the input.
This tests whether the inferred event state describes the shower response beyond the observed locations.
Such response completion may be useful when individual stations fail or only sparse measurements are available.
We compare both ordinary reconstruction and super-resolution with the standard TA reconstruction procedure.

Our primary goal is to learn compact event embeddings from measurements and use them to reconstruct and analyze events.
The IceCube photon-transport surrogate takes specified source properties and sensor positions as input~\cite{Yu2026Candela} to predict the outputs.
In our approach, the encoder infers an event description from its measurements, and the shared MLP connects this description to the predicted detector response.
The same embedding supports reconstruction, recovery of missing measurements, and comparisons between events in anomaly searches.
Parameter conditioning and established adversarial regularization allow the scientific question to guide which variations define an anomaly.

\section{Method}
\label{sec:arhitecture}

\subsection{Telescope Array surface detector and event data}

\begin{figure}[htbp]
    \centering
    \begin{minipage}[b]{0.475\textwidth}
        \centering
        \includegraphics[width=\linewidth]{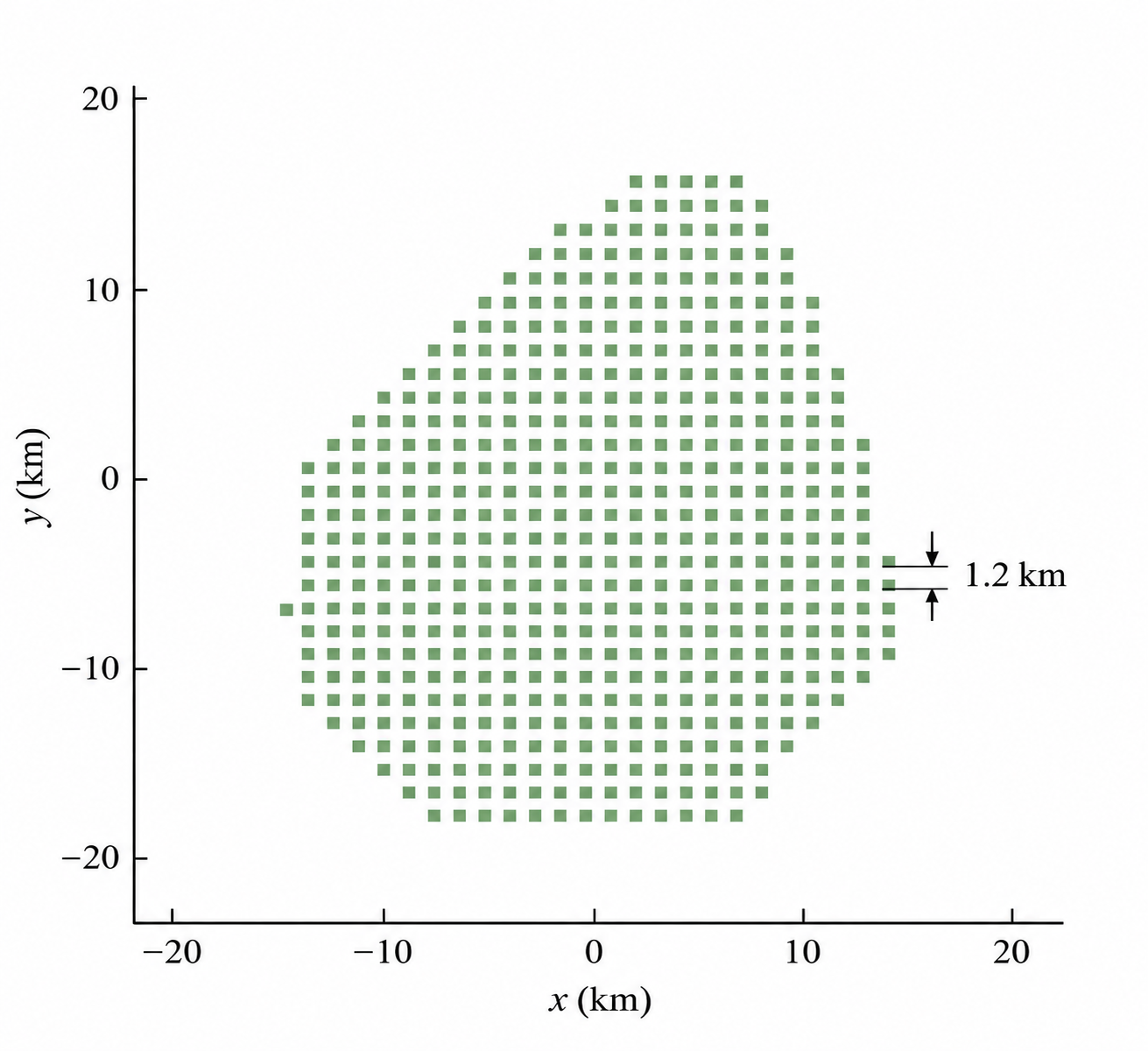}
        \caption{
            Layout of the Telescope Array detectors.
            Squares denote the 507 surface detector stations.
        }
        \label{fig:TA_layout}
    \end{minipage}
    \hfill
    \begin{minipage}[b]{0.475\textwidth}
        \centering
        \raisebox{0.6cm}{%
            \includegraphics[width=\linewidth]{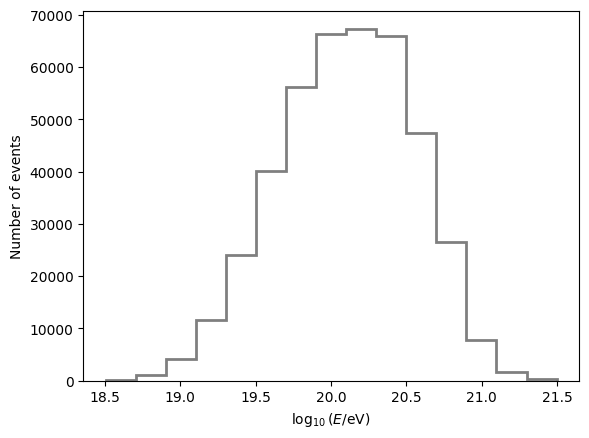}%
        }
        \caption{
            Primary-energy distributions of the simulated proton and photon samples.
        }
        \label{fig:energy_spectrum}
    \end{minipage}
\end{figure}

The Telescope Array (TA) experiment is located in Utah, USA, and studies ultra-high-energy cosmic rays~\cite{AbuZayyad2012}.
Its measurements of their energy spectrum, arrival directions, and mass composition aim to clarify their origin and propagation~\cite{AbuZayyad2013Spectrum,Abbasi2014Hotspot,Abbasi2018Xmax}.
The surface detector records extensive air showers with an array of 507 scintillation stations.
The stations are arranged on a regular grid with a spacing of approximately \(1.2\,\mathrm{km}\), covering about \(700\,\mathrm{km}^2\).
Each station contains two layers of plastic scintillator.
The analysis uses the integrated station signal and the arrival time of the shower front.
The detector layout is shown in Fig.~\ref{fig:TA_layout}.

The surface array samples the particles of an extensive air shower (EAS) at ground level.
Only a subset of the array contributes, so the number of stations varies between events.
The integrated station signals describe the shower's spatial footprint, while the station activation times trace the passage of the shower front.
Together, these measurements contain information about shower development and arrival direction.
For each triggered station, we use its three spatial coordinates, integrated signal, and activation time.
Signals are measured in minimum-ionizing-particle (MIP) units, with \(1\,\mathrm{MIP}\approx2.05\,\mathrm{MeV}\) for TA~\cite{AbuZayyad2012}.
An event consists of the responses of the stations triggered by a single shower.

The station coordinates \(x\) and \(y\) are measured relative to the reconstructed shower core.
The core is the point where the reconstructed shower axis intersects the ground-level surface~\cite{Ivanov2021}.
The \(z\) coordinate corresponds to the station height relative to a fixed reference level of the experiment.
Station activation times are measured relative to the reconstructed time at which the shower front reaches the core~\cite{Ivanov2021}.
This provides a common position and time reference for comparing shower footprints.
Each triggered station is therefore represented by the vector:
\begin{equation}
\mathbf{x}_i = \left(\mathbf{r}_i, Q_i, t_i\right)
\end{equation}
where \(\mathbf{r}_i=(x_i,y_i,z_i)\) denotes the station coordinates, \(Q_i\) is the integrated signal, and \(t_i\) is the relative station trigger time.

Monte Carlo simulated proton- and photon-induced showers are used for model training and evaluation.
Shower development is simulated using CORSIKA~\cite{Heck1998} with the QGSJET-II-04 hadronic interaction model~\cite{Ostapchenko2011}.
The primary-energy distributions of the simulated samples are shown in Fig.~\ref{fig:energy_spectrum}.
Proton and photon samples have identical primary-energy distributions.
This avoids introducing a trivial spectral difference into the anomaly-search benchmark.
The proton sample contains approximately \(1.90\times10^{6}\) events.
The data are randomly divided into training and test subsets in a \(9{:}1\) ratio.
All input features are normalized using statistics from the training dataset.

\subsection{Autoencoder}

\begin{figure}[htbp]
    \centering
    \includegraphics[width=0.95\textwidth]{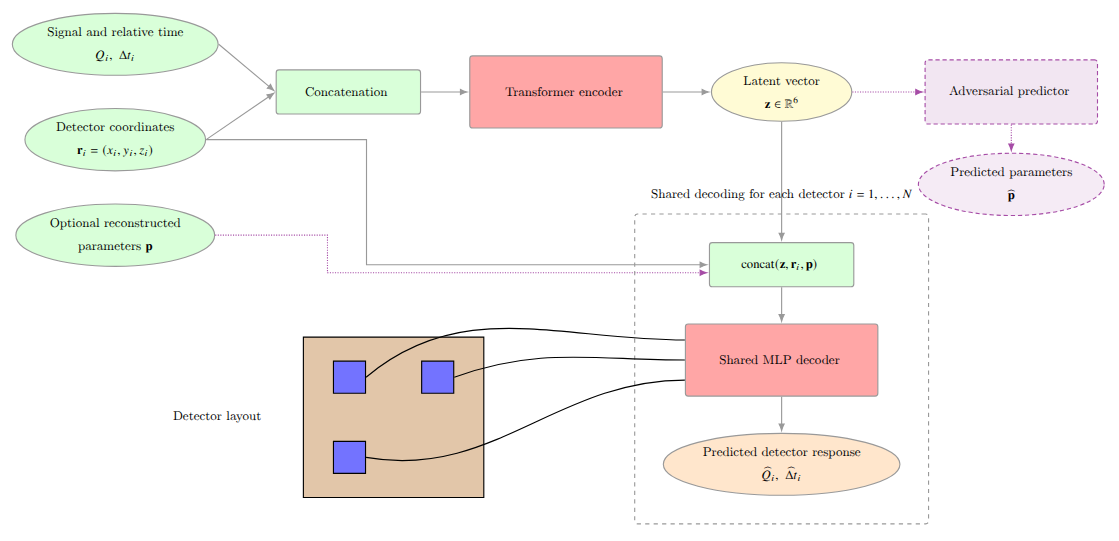}

    \caption{
Schematic of the developed autoencoder architecture.
The Transformer encoder maps the set of features from triggered detector stations to a six-dimensional latent representation of the event. The same fully connected decoder is applied to each station and reconstructs its integrated signal and relative activation time from the latent vector, the station coordinates, and, when required, additional reconstructed parameters. The optional adversarial block attempts to reconstruct selected physical parameters from the latent vector, while the encoder is trained to hinder such prediction.
    }
    \label{fig:model_architecture}
\end{figure}

We construct a geometry-conditioned autoencoder that learns a compact event description by reconstructing the detector measurements.
The general architecture is shown in Fig.~\ref{fig:model_architecture}.
An encoder maps the available station measurements to a common latent vector \(\mathbf{z}\).
A shared decoder uses this vector and the station coordinates to reconstruct integrated signals and activation times.
To get the full TA detector response, the MLP decoder is queried at each station position.
The encoder and decoder are trained jointly to obtain an event embedding that supports both response reconstruction and subsequent analysis.

The encoder, which is a four-layer Transformer-encoder~\cite{Vaswani2017}, receives the feature vectors \(\mathbf{x}_i\) of the \(N\) triggered stations in an event.
Through self-attention, the encoder learns useful correlations between station signals and relative activation times at different positions.
It can therefore use event-wide patterns while accepting different numbers of stations.
The resulting common event vector is
\begin{equation}
\mathbf{z}
=
E\left(\{\mathbf{x}_i\}_{i=1}^{N}\right)
\in\mathbb{R}^{6}.
\end{equation}
We use six latent components.
A smaller latent space substantially degraded reconstruction in our tests.
The vector describes the shower as a whole, rather than an individual station.

To turn the event description into local detector responses, we use a shared MLP decoder.
For a station at coordinates \(\mathbf{r}_i\), it predicts the integrated signal and activation time:
\begin{equation}
D\left(\mathbf{z},\mathbf{r}_i\right)  
=
\left(\hat{Q}_i,\hat{t}_i\right).
\end{equation}
To reconstruct the full event, the decoder is applied sequentially to all considered stations using the same latent vector \(\mathbf{z}\).

For a fixed latent vector, the MLP defines continuous fields of integrated signal and relative activation time, acting as a neural-field surrogate of the shower response~\cite{Mildenhall2020}.
Shower energy, arrival direction, and the primary particle affect the spatial signal pattern and the shape of the shower front.
The decoder learns how their imprint in the common event description translates into a local detector response.
Physically, it generalizes the lateral distribution and shower-front functions used in standard TA reconstruction~\cite{Ivanov2021}.
Rather than fixing their analytical form, we learn the response family from simulated showers.
This gives the model the flexibility to describe variations in shower development and their effects across the array.

We train this event representation using the InfoVAE formulation with maximum mean discrepancy (MMD) regularization~\cite{Zhao2017InfoVAE}.
The training objective is
\begin{equation}
\mathcal{L}_{\mathrm{VAE}}
=
\mathcal{L}_{\mathrm{rec}}
+
\lambda_{\mathrm{MMD}}\mathcal{L}_{\mathrm{MMD}},
\end{equation}
where \(\lambda_{\mathrm{MMD}}=0.05\).
In our experiments, a conventional VAE~\cite{Kingma2013} showed posterior collapse, hence the KL-divergence term is not used.
The reconstruction loss is the mean squared error of the integrated signals and activation times, evaluated only at triggered stations.
The MMD term penalizes differences between the distribution of embeddings across events and the chosen normal prior.

In our comparison with LSTM~\cite{Hochreiter1997} and Transformer decoders, the MLP gave a lower mean reconstruction error with fewer trainable parameters.
This architecture assigns the MLP a compact physical task: learning the shared response field with the detector geometry supplied explicitly.

\subsection{Conditional transfer of physical parameters}

Some physical parameters affect the detector response but need not define the anomaly being sought.
For example, we may seek showers with unusual structure at a given energy.
A rare energy alone should not determine the anomaly score in such a search.
We therefore allow selected parameters to enter the decoder as known physical conditions.

Let \(\mathbf{p}\) denote these parameters.
The conditioned decoder reconstructs the integrated signal and relative activation time as
\begin{equation}
D\left(\mathbf{z},\mathbf{r}_i,\mathbf{p}\right)
=
\left(\hat{Q}_i,\hat{t}_i\right).
\end{equation}
We demonstrate this procedure using reconstructed primary-particle energy as the conditioning parameter.
Arrival direction and other reconstructed quantities could be treated in the same way.

Providing \(\mathbf{p}\) to the decoder does not by itself prevent the encoder from storing the same information in \(\mathbf{z}\).
To discourage this redundancy, we use an established adversarial representation-learning method~\cite{Wang2018Invariant}.
An MLP critic predicts the supplied parameters from the event embedding, \(\hat{\mathbf{p}}=C(\mathbf{z})\).
Its prediction loss is
\begin{equation}
\mathcal{L}_{\mathrm{adv}}
=
\left\|\mathbf{p}-\hat{\mathbf{p}}\right\|_2^2.
\end{equation}

The critic minimizes \(\mathcal{L}_{\mathrm{adv}}\), while the encoder is trained to oppose this prediction through the autoencoder objective
\begin{equation}
\mathcal{L}_{\mathrm{AE}}
=
\mathcal{L}_{\mathrm{VAE}}
-
\lambda_{\mathrm{adv}}\mathcal{L}_{\mathrm{adv}},
\end{equation}
where \(\lambda_{\mathrm{adv}}=0.1\).
The reconstruction loss requires the encoder to preserve useful shower information, while the adversarial term discourages information already supplied through \(\mathbf{p}\).

The conditioning input lets the decoder retain the physical energy dependence of the station responses.
The embedding is instead encouraged to describe shower structure beyond this known energy.
We evaluate the remaining dependence through the energy variation of the anomaly score and the ability to recover energy from the embedding.

The autoencoder is trained for 100 epochs using AdamW~\cite{Loshchilov2019}, with a batch size of 200.
The initial learning rate is \(5\times10^{-5}\).
We use ReduceLROnPlateau with a reduction factor of 0.2 and a patience value of 5.
The encoder and decoder contain approximately \(153\,000\) and \(20\,000\) trainable parameters, respectively.
During training, Gaussian noise with a standard deviation of 0.05 is added to the integrated-signal and relative activation-time channels.

\section{Results}
\label{sec:result}

\subsection{Detector-wise generation}

We first test how accurately the autoencoder reproduces the station responses of an event.
We compare the predicted integrated signals and relative activation times with those from the standard TA reconstruction~\cite{Ivanov2021}.
The latter uses analytical parametrizations of the lateral signal distribution and shower front.
The comparison uses the root mean squared error (RMSE) of the observables in physical units (MIP for signals and \(\mu\mathrm{s}\) for activation times).

\begin{figure}[htbp]
    \centering
    \includegraphics[
        width=\textwidth
    ]{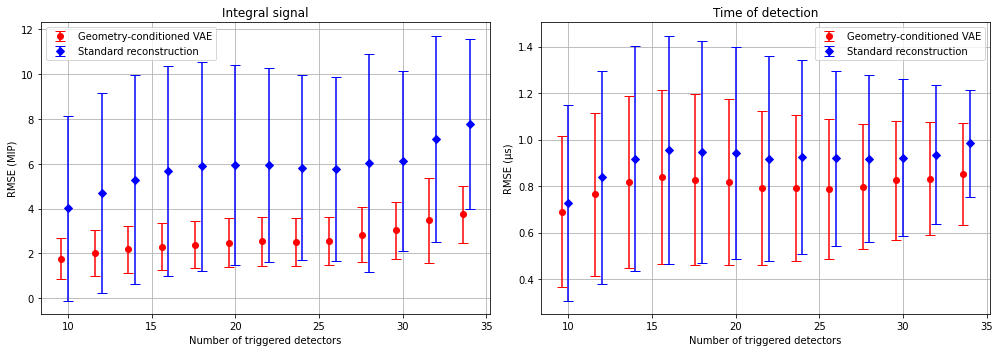}
    \caption{
        Dependence of the root mean squared reconstruction error
        on the number of triggered detectors for the standard
        procedure (\textit{Standard reconstruction}) and the proposed model
        (\textit{Geometry-conditioned VAE}).
        The error is given in MIP for the integrated signal (left) and in \(\mu\mathrm{s}\) for the relative activation time (right).
        Points show the mean event-wise RMSE over all events with the same
        number of triggered stations; error bars indicate the standard
        deviation of the event-wise RMSE within each group.
    }
    \label{fig:STD_ALL}
\end{figure}

Figure~\ref{fig:STD_ALL} shows the reconstruction error as a function of the number of triggered stations.
The autoencoder outperforms the standard reconstruction in predicting integrated signals and activation times in all of the investigated range.
The smaller gain for timing may reflect that shower-front curvature is easier to parametrize analytically than the lateral signal distribution. 
The shared MLP therefore learns a more precise shower-response mapping than the analytical approximation used in the standard reconstruction.
For any encoded event, the surrogate can be queried at arbitrary positions.
This gives the full lateral distribution function and relative activation-time profile.

\subsection{Super-resolution}

Detector failures and temporary interruptions in data taking can leave gaps in the measured shower profile.
A response surrogate can fill these gaps by estimating the missing integrated signals and relative activation times from the working stations.
We test this capability by withholding station measurements from the encoder and reconstructing the original event.

For this test, we retain a subgrid with twice the original spacing in each horizontal direction.
The station closest to the reconstructed shower axis is also retained.
The encoder thus receives approximately four times fewer stations than the decoder must reconstruct.
The decoder predicts integrated signals and relative activation times at all station positions in the original event.
Only events with at least four triggered stations after this selection are included.

\begin{figure}[htbp]
    \centering
    \includegraphics[
        width=\textwidth
    ]{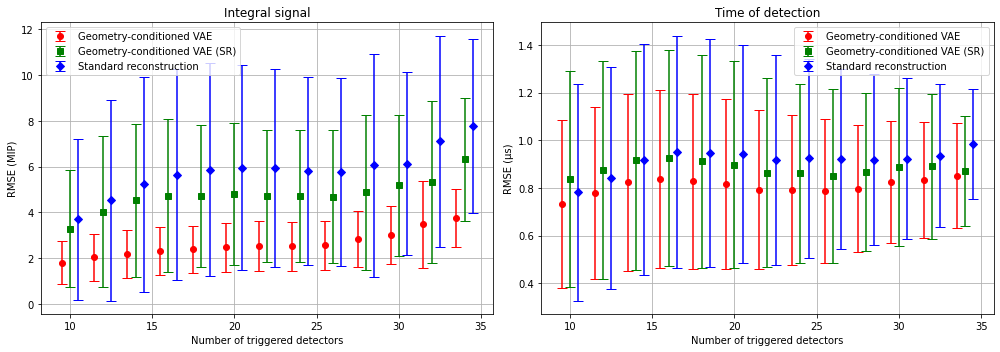}
    \caption{
        Dependence of the root mean squared reconstruction error
        on the number of detectors for the super-resolution task when
        reconstructing the full detector grid from a thinned input event.
        \textit{Geometry-conditioned VAE} denotes the model
        in the normal reconstruction mode,
        whereas \textit{Geometry-conditioned VAE SR} denotes the model
        in the super-resolution mode.
        The error is given in MIP for the integrated signal (left) and in \(\mu\mathrm{s}\) for the relative activation time (right).
        Points show the mean event-wise RMSE over all events with the same
        number of triggered stations; error bars indicate the standard
        deviation of the event-wise RMSE within each group.
    }
    \label{fig:STD_SR}
\end{figure}

\begin{figure}[htbp]
    \centering
    \includegraphics[
        width=\textwidth
    ]{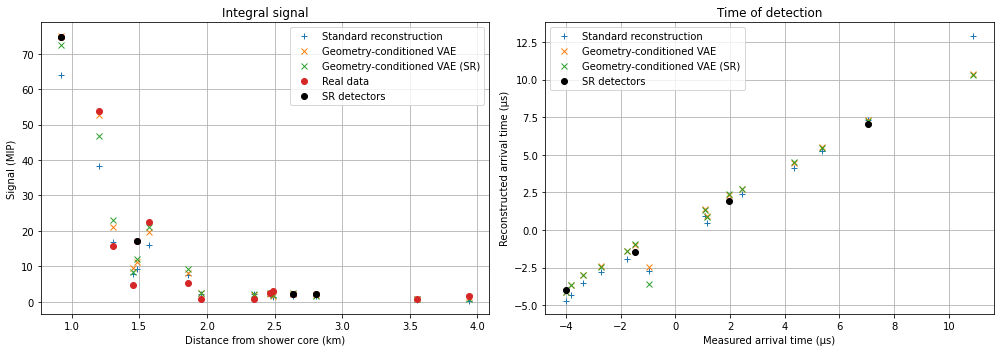}
    \caption{
        Example reconstruction of a single event in the normal autoencoder mode
        (\textit{Geometry-conditioned VAE}) and in the super-resolution mode
        (\textit{Geometry-conditioned VAE SR}) compared with the standard reconstruction.
        Left: integrated signal as a function of the distance to the shower core;
        right: reconstructed detector trigger time compared with the true value.
    }
    \label{fig:reconstruction_example}
\end{figure}

Figure~\ref{fig:STD_SR} shows that reducing the number of input stations increases the reconstruction error relative to the full-input model.
The reduced-input model nevertheless predicts station responses more accurately than the standard TA procedure.
Figure~\ref{fig:reconstruction_example} shows an example of per-station reconstruction for a single event in the full-input and super-resolution modes.
The high reconstruction accuracy at withheld stations suggests that shower information and detector geometry are not entangled in the latent representation.
This permits interpolation of the integrated signals and activation times between stations, enabling super-resolution.

% TODO decrease the vertical size of the figure

The increased reconstruction error is expected because fewer station measurements constrain the encoder, leaving less information about the shower in the latent vector.
Excluding events with fewer than four retained stations also reduces the training sample and may contribute to the error increase.
The decoder can still be queried at arbitrary positions where the Monte Carlo response is reliable.
The accuracy of these predictions, however, remains limited by the information in the original sparse measurements.

\subsection{Anomaly measure}

We evaluate the learned event representation for identifying showers with unusual structure.
As a case study, we consider photon-induced showers in a proton-dominated background as a controlled benchmark. 
Everywhere below --- for model training and anomaly search --- we use the mixed sample mimicking the real data analysis case, when the experimental sample has all types of events.
This avoids assuming access to an anomaly-free reference population in an experimental search.
The photon fraction in our dataset is \(0.8\%\).

A CatBoost classifier~\cite{Prokhorenkova2018} was trained on the latent  vectors for supervised proton--photon classification. 
Selecting events above the 85th percentile of the classifier score retains 84.8\% of  photons and rejects 85.6\% of protons.
This demonstrates that the embedding retains information about the primary species.
The anomaly search tests whether that information can be used to select rare showers without a classifier trained on primary-particle labels.

Note that a shower can be reconstructed accurately and still have an atypical event embedding.
We therefore combine the compact event description with the reconstruction error when estimating event rarity.
Thus, each event is represented by the six latent components and the event reconstruction error:
\begin{equation*}
\mathbf{v}=(\mathbf{z},L_{\mathrm{rec}})\in\mathbb{R}^{7}.
\end{equation*}
We fit a Gaussian kernel density estimate~\cite{Parzen1962} with bandwidth \(h=0.3\) to this joint representation.
The bandwidth was selected empirically to avoid an excessively peaked density estimate, with high values near individual reference events and nearly zero density between them.

The anomaly score is
\begin{equation*}
A(\mathbf{v})=-\log\hat{p}(\mathbf{v}).
\end{equation*}
It ranks events by their rarity in the reference population, without defining a photon class.

\begin{figure}[htbp]
    \centering
    \includegraphics[width=0.65\textwidth]{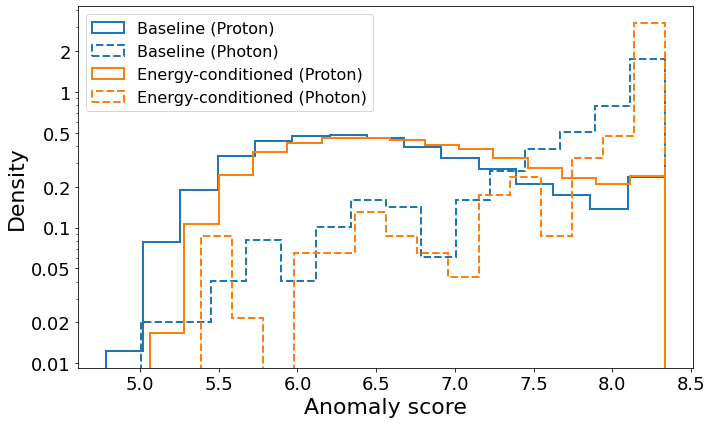}
    \caption{
        Distributions of the anomaly score
        for proton and photon events.
        The label \textit{Baseline} corresponds to the standard architecture,
        whereas \textit{Energy-conditioned} corresponds to the model
        with energy provided to the decoder.
    }
    \label{fig:KDE}
\end{figure}

To reduce the influence of the discrete station multiplicity, we fit the KDE to events with \(N=15\text{--}18\) triggered stations.
Selecting events above the 85th percentile of the score retains \(73.3\%\) of photons and rejects \(85.5\%\) of protons.
The baseline distribution in Fig.~\ref{fig:KDE} shows the resulting concentration of photon events in the high-score tail.
Comparable selection performance is also obtained with broader station-multiplicity ranges.

We compare the joint score with reconstruction error alone~\cite{Finke2021} and with KDE in the six-dimensional latent space.
At the same percentile threshold, reconstruction error alone retains \(57.5\%\) of photons and rejects \(85.6\%\) of protons.
Although photon showers tend to have larger errors, the broad error distribution limits this selection.
KDE applied only to the six latent components selects \(27.5\%\) of photons.
The improvement from combining the two quantities indicates that the embedding and the reconstruction residual provide complementary information about atypical showers.
Intersecting the latent-only KDE and reconstruction-error selections would retain at most \(27.5\%\) of photons at these thresholds.
We therefore include the reconstruction error in the KDE input rather than impose both cuts.

The anomaly score remains dependent on the population used for density estimation.
Increasing the photon fraction makes photon showers less rare and reduces their detection efficiency.
Selected events may therefore reflect genuine physical anomalies or detector malfunctions.
The defining criterion is the rarity of the measured response in the reference population, not membership of a predefined particle class.

\subsection{Anomaly measure with conditional transfer of physical parameters}

The choice of which physical parameters should define an anomaly depends on the search.
Here we target unusual shower development at a known reconstructed energy.
We therefore treat energy as a condition and seek to reduce its contribution to the anomaly score.

The baseline is the geometry-conditioned autoencoder used in the preceding subsection.
The energy-conditioned model supplies reconstructed energy directly to the decoder and adds a critic that discourages its encoding in the latent vector.

The photon benchmark tests whether this constraint preserves sensitivity to unusual shower structure.
On the same sample and at the same 85th-percentile threshold, photon retention rises from \(73.3\%\) to \(79.3\%\), while proton rejection remains \(85.5\%\).
Figure~\ref{fig:KDE} shows the corresponding improvement in the separation of photon events from the proton background.

\begin{figure}[htbp]
    \centering
    \includegraphics[width=0.7\textwidth]{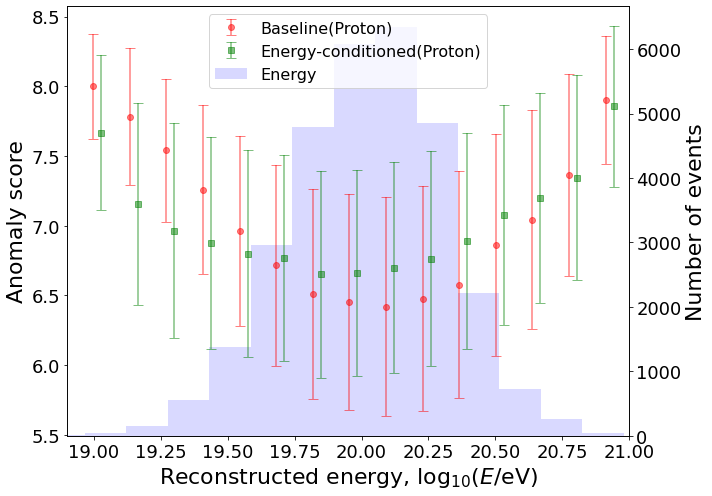}
    \caption{
        Dependence of the anomaly score on the reconstructed
        primary-particle energy for proton events.
        The baseline model and the model with conditional transfer
        of energy directly to the decoder are shown.
        The background histogram shows the distribution of events
        over energy.
        Points show the mean anomaly score over proton events within each
        reconstructed-energy bin; error bars indicate the standard deviation
        of the anomaly score within each bin.
    }
    \label{fig:anomaly_std}
\end{figure}

The effect on proton events clarifies the role of the energy conditioning.
In the baseline model, Fig.~\ref{fig:anomaly_std} shows elevated scores in the sparsely populated tails of the energy distribution.
Energy conditioning weakens this dependence --- the Spearman rank correlation between reconstructed energy and anomaly score changes from \(-0.42\) to \(-0.14\).

We also test the energy information in the latent vector with an auxiliary regression model.
Its mean squared error increases from \(0.28\) for the baseline to \(0.50\) for the energy-conditioned model.
The energy-conditioned latent space therefore contains less information about energy, and the energy is passed to the decoder as an explicit additional input to the station-response model.

\section{Discussion}
\label{sec:discussion}

The present study demonstrates detector-response reconstruction, super-resolution, and unsupervised anomaly detection on simulated Telescope Array data.
We now discuss possible applications to event-parameter reconstruction, calibration with experimental data, and fast Monte Carlo generation.

The conditional decoder can serve as a forward response model for primary-particle energy reconstruction.
Established approaches such as IceCube Event-Generator fit event parameters using learned response likelihoods~\cite{Huennefeld2021EventGenerator}.
A suitable uncertainty model for the signal and timing residuals can define the reconstruction likelihood for our surrogate.
Specifically, the encoder first determines the event embedding of the measured shower without an explicit energy input.
With this embedding fixed, gradient descent can minimize the negative log-likelihood over the decoder's energy input to find the best reconstruction of the shower.

One of the key features of surrogate models is that they can be calibrated directly against experimental data.
For example, neural surrogates have been developed for tuning JUNO's simulated energy response~\cite{Gavrikov2026Calibration}.
Cross-calibration against an independent energy reference is standard experimental practice, used in both Pierre Auger~\cite{Aab2020Spectrum} and TA~\cite{AbuZayyad2013Spectrum}.
Hybrid events link the measured SD response to the calorimetric energy estimate from the fluorescence detector (FD).

We propose the following calibration scheme on the TA example.
A trained encoder first obtains the event vector from the measured SD station data.
The FD energy then replaces the reconstructed energy supplied to the conditional decoder.
With the encoder frozen, only the shared MLP is fine-tuned to reproduce the measured integrated signals and relative activation times at the station coordinates.
With this calibration, fitting the energy supplied to the decoder directly estimates the primary-particle energy on the FD scale from SD measurements alone.

Finally, the MLP decoder can also be used for fast MC within established generative autoencoder methods such as BIB-AE~\cite{Buhmann2021BIBAE}.
In this scheme, a latent sample specifies a shower state, and the energy-conditioned MLP evaluates the response at the chosen station positions.
In our decoder comparison, the shared MLP gave a lower mean reconstruction error than LSTM and Transformer decoders while using fewer trainable parameters.

\section{Conclusions}
\label{sec:conclusion}

We have developed a geometry-conditioned autoencoder that separates the physics of the extensive air shower, encoded in the latent space, from detector geometry.
The shared MLP acts as a neural-field response surrogate that learns the physical shower response, while station coordinates specify where it is evaluated.

In the TA Monte Carlo benchmark, this learned response mapping predicts station observables more accurately than the standard analytical reconstruction.
With inputs thinned to twice the original spacing, it recovers responses across the original \(1.2\,\mathrm{km}\) station grid from approximately four times fewer measurements.
The inferred event description thus supports continuous signal and timing profiles, including estimates at non-working stations, with precision limited by the available measurements.

The joint density of the event vector and reconstruction error defines a target-agnostic anomaly score.
Including the event vector allows a well-reconstructed shower to be selected when its representation is rare.
Supplying selected physical parameters to the decoder and suppressing their latent encoding with a critic lets the scientific objective determine the anomaly criterion.
For energy conditioning, this targets unusual shower development at a known energy rather than spectral rarity.

With energy treated as known, the score retains \(79.3\%\) of photon events and rejects \(85.5\%\) of protons.
This increases the photon fraction in the selected sample by approximately a factor of \(5\).
While a specialized photon search can use labeled examples of a known target class to achieve stronger separation, our method provides a general unsupervised search without a predefined anomalous class.

%%%%%%%%%%%%%%%%%%%%
\section*{Acknowledgments}
%%%%%%%%%%%%%%%%%%%%
This work was supported by the Russian Science Foundation under grant no. 25-12-00333.

%%%%%%%%%%%%%%%%%%%%
\section*{Code availability}
%%%%%%%%%%%%%%%%%%%%

The code for reproducing the results of the paper is available at {\small\url{https://github.com/fitagdinov/geometry-conditioned-autoencoders}}.

\bibliography{references}

\end{document}